# Shape-Dependent Multi-Weight Magnetic Artificial Synapses for Neuromorphic Computing

Thomas Leonard[1], Samuel Liu[1], Mahshid Alamdar[1], Can Cui[1], Otitoaleke G. Akinola[1], Lin Xue[2], T. Patrick Xiao[3], Joseph S. Friedman[4], Matthew J. Marinella[3], Christopher H. Bennett*[3] and Jean Anne C. Incorvia*[1]

[1]*Electrical and Computer Engineering Dept., University of Texas at Austin, Austin, TX, USA.*

[2]*Applied Materials, Santa Clara, CA, USA.*

[3]*Sandia National Laboratories, Albuquerque, NM, USA.*

[4]*Electrical and Computer Engineering Dept., University of Texas at Dallas, Richardson, TX, USA.*

In neuromorphic computing, artificial synapses provide a multi-weight conductance state that is set based on inputs from neurons, analogous to the brain. Additional properties of the synapse beyond multiple weights can be needed, and can depend on the application, requiring the need for generating different synapse behaviors from the same materials. Here, we measure artificial synapses based on magnetic materials that use a magnetic tunnel junction and a magnetic domain wall. By fabricating lithographic notches in a domain wall track underneath a single magnetic tunnel junction, we achieve 4-5 stable resistance states that can be repeatably controlled electrically using spin orbit torque. We analyze the effect of geometry on the synapse behavior, showing that a trapezoidal device has asymmetric weight updates with high controllability, while a straight device has higher stochasticity, but with stable resistance levels. The device data is input into neuromorphic computing simulators to show the usefulness of application-specific synaptic functions. Implementing an artificial neural network applied on streamed Fashion-MNIST data, we show that the trapezoidal magnetic synapse can be used as a metaplastic function for efficient online learning. Implementing a convolutional neural network for CIFAR-100 image recognition, we show that the straight magnetic synapse achieves near-ideal inference accuracy, due to the stability of its resistance levels. This work shows multi-weight magnetic synapses are a feasible technology for neuromorphic computing and provides design guidelines for emerging artificial synapse technologies.

INTRODUCTION

The human brain is incredibly efficient at processing unstructured information in real time. In the brain, synapses control the degree of connectivity between neurons, and this has been analogously implemented by synapses in artificial neural networks (NNs). An artificial synapse implemented in an analog electrical device has multi-weight (MW) conductance states that can be set based on inputs from neurons. Emerging understanding of the brain shows that the synaptic connection has many useful properties beyond MW to aid in learning[1,2]. Analogous properties in artificial synapses have been shown to be beneficial in NNs, but the desired properties of the synapse depend on the application. One desired property is *controllability*, i.e. a given input sets the synapse to a given state (weight in context of a NN), while still having a certain degree of *stochasticity*[3–5] to avoid local minima. Neural networks that behave probabilistically are also increasingly used to emulate the stochastic spiking behavior of neurons, as well as to quantify the network's confidence in its predictions[6]. Some applications, such as online training using backpropagation, have additional requirements such as *symmetry*, i.e. the MW has equal but opposite responses to positive and negative electrical stimuli[7]. By contrast, single-shot programming[8], where a given voltage can bring the synapse directly to a given weight, benefits from *asymmetry*, i.e. the various weight states can be set in a positive voltage direction and fully reset in the opposite.

While a number of materials classes are being used as synapses, including resistive random-access memory (RRAM)[9] and ferroelectric memory (Fe-RAM)[10], spintronic solutions benefit from being naturally stochastic[11], and magnetic random-access memory (MRAM) has been shown to have high endurance[12], scalability down to 25 nm feature sizes[13], and CMOS compatibility[12]. To achieve a number of conductance states per device, this paper focuses on the three-terminal domain wall-magnetic tunnel junction (DW-MTJ) device, where a magnetic domain wall (DW) is electrically pushed back and forth underneath a magnetic tunnel junction (MTJ), changing the MTJ's conductance. For neuromorphic computing, the DW-MTJ may offer additional benefits, including the realization of various bio-mimetic functions in the same film stack through modifications to the device geometry. For example, the same film stack can behave both as a synapse or as a leaky, integrate, and fire neuron with intrinsic leaking[14–16,17,18]. The DW-MTJ can also have device-to-device magnetic field interactions to mimic the connectivity of the brain[19], and it can be used in both artificial neural networks (ANNs)[14] and spiking neural networks (SNNs)[15]. When used for online learning with ANNs, we have previously used micromagnetic simulations to prove that the DW-MTJ synapse is sufficiently linear, controllable, and possesses low enough device variation to make the lower on/off ratio (expressed as tunnel magnetoresistance, or $TMR$) of the sensing MTJ tolerable[7]. However, experimental work has focused mostly on binary DW-MTJs[20,21], with only one MW MTJ paper demonstrating a multi-MTJ device[22] that could have scalability challenges. Several works have investigated theoretical and experimental stochastic functionality of magnetic devices[6,23–25], but none have also included MW switching. A demonstration of both controllability and stochasticity of a MW DW-MTJ synapse is a critical step in order to suggest a new species of multifunctional synapse.

Here, we demonstrate DW-MTJ MW synapses that use a single MTJ for read-out and achieve notably low variation in the programmed resistance between write cycles. We analyze their controllability vs. stochasticity, geometry effects, and neuromorphic applications. Lithographic notches are used to set the DW at specific and reproducible positions along a magnetic DW track. Two device geometries with different synaptic behavior profiles, but identical material stacks, are demonstrated: a trapezoidal geometry and a straight geometry. Spin orbit torque (SOT) is used to drive the DW between notches underneath the MTJ, and the resulting synapse state is measured by reading the resistance of the MTJ. We show 4-5 stable resistance states with a variation in switching voltage as low as 2.5% over ten cycles, and variation in resistance at a given weight as low as $\pm 0.3$ Ω. We then simulate NN tasks using the device data. We show that the trapezoidal geometry can emulate a metaplastic function that allows an ANN to learn as Fashion-MNIST data is presented (stream learning), which is useful for edge computing, where a computer should continue to learn while immersed in its environment. We show that the straight geometry can achieve near-ideal inference accuracy, due to the high stability of its resistance levels, by simulating a convolutional NN for CIFAR-100 image recognition. The DW-MTJ synapse device metrics are compared to other state-of-the-art synapse technologies. This work shows MW magnetic synapses are a feasible technology for neuromorphic computing while also providing design guidelines for emerging synapse technologies.

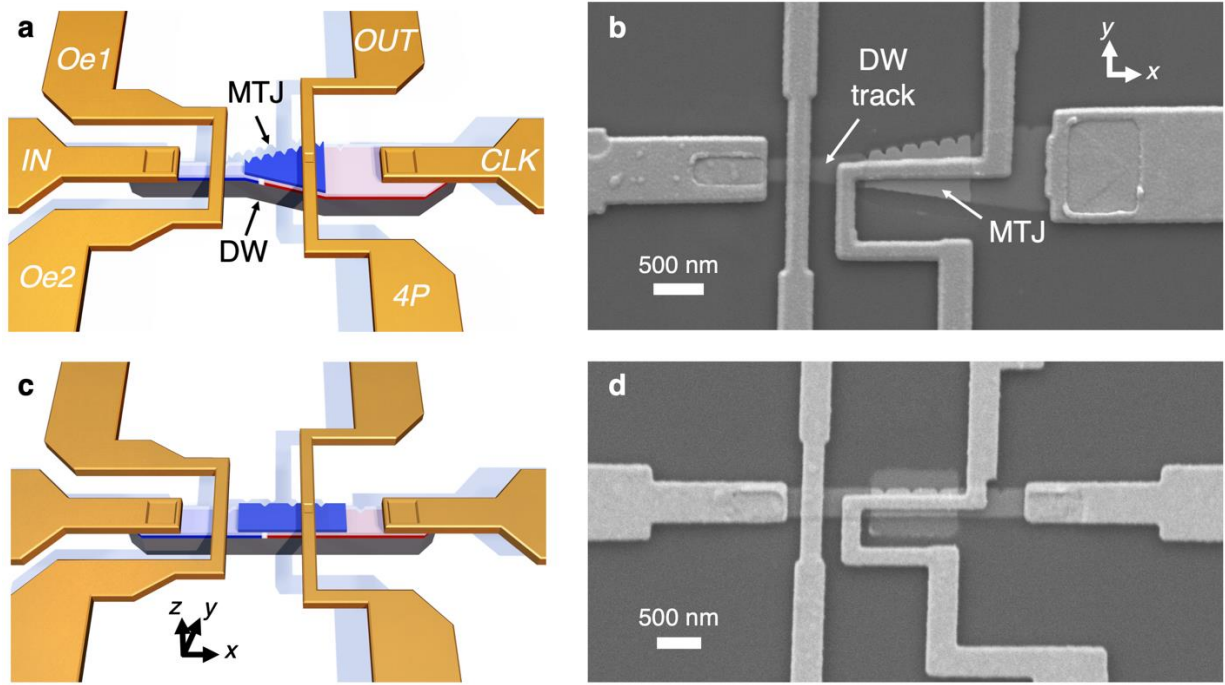

**Figure 1. DW-MTJ synapse device structure. a**, DW-MTJ synapse cartoon with trapezoidal DW track. Ta (grey)/CoFeB (red and blue)/MgO (white) DW track with white DW and lithographic notches is shown. MTJ output is depicted in blue. **b**, Top-down SEM of the DW-MTJ trapezoidal synapse with 250 nm minimum DW track width. **c**, DW-MTJ synapse cartoon with straight DW track. **d**, Top-down SEM of the DW-MTJ straight synapse with 350 nm DW track width.

# ELECTRICAL CHARACTERIZATION OF MAGNETIC SYNAPSE PROTOTYPES

To fabricate the devices, following the material and processing details of Reference [20], an MTJ film stack with perpendicular magnetic anisotropy was grown and then patterned using electron beam lithography and ion beam etching (see *Methods*). The device includes a bottom free magnetic switching layer with an underlayer Tantalum heavy metal for SOT-driven DW motion, and a pinned top reference magnetic layer. A device cartoon and corresponding top-down scanning electron microscope (SEM) image for the two geometries studied are shown in Fig. 1. The MTJ extends along the DW track and pinning notches were fabricated. The *IN* and *CLK* terminals are used to drive the DW back and forth using SOT; the $Oe_1$ and $Oe_2$ terminals are used to electrically initialize the DW position; and the *4P* and *OUT* terminals are used to measure the MW resistance state $R_{MTJ}$. In Fig. 1a-b, the DW track was fabricated in a trapezoidal shape with one edge 4× wider than the other, while in Fig. 1c-d the track is a straight rectangle.

Figure 2a shows the out-of-plane minor field loop of the trapezoidal synapse, showing perpendicular magnetic anisotropy of the free layer and an offset field loop due to stray field from the pinned layers. The $TMR = \frac{R_{MTJ,max} - R_{MTJ,min}}{R_{MTJ,min}} = 54\%$ with resistance-area product $RA = R_{MTJ,min} \times MTJ\ area = 60\ \Omega-\mu m^2$ and 33 Ω difference between the maximum and minimum MTJ resistances.

SOT switching opens up additional resistance states at the lithographically-defined pinning locations under the MTJ. The device is first set to the antiparallel configuration using a DC 50 mT ($\hat{z}$) saturating magnetic field (the free and reference layers have antiparallel magnetization). After this step, a DC 4 mT ($-\hat{z}$) magnetic field is applied to compensate for the offset field loop and promote efficient SOT switching, which could be removed through film stack engineering. Then, the write is performed when a 50 nanosecond (ns) voltage pulse (10 ns rise and fall times) with voltage amplitude $V$ is applied between the *IN* and *CLK* terminals with increasing amplitude. After each pulse, $R_{MTJ}$ is measured.

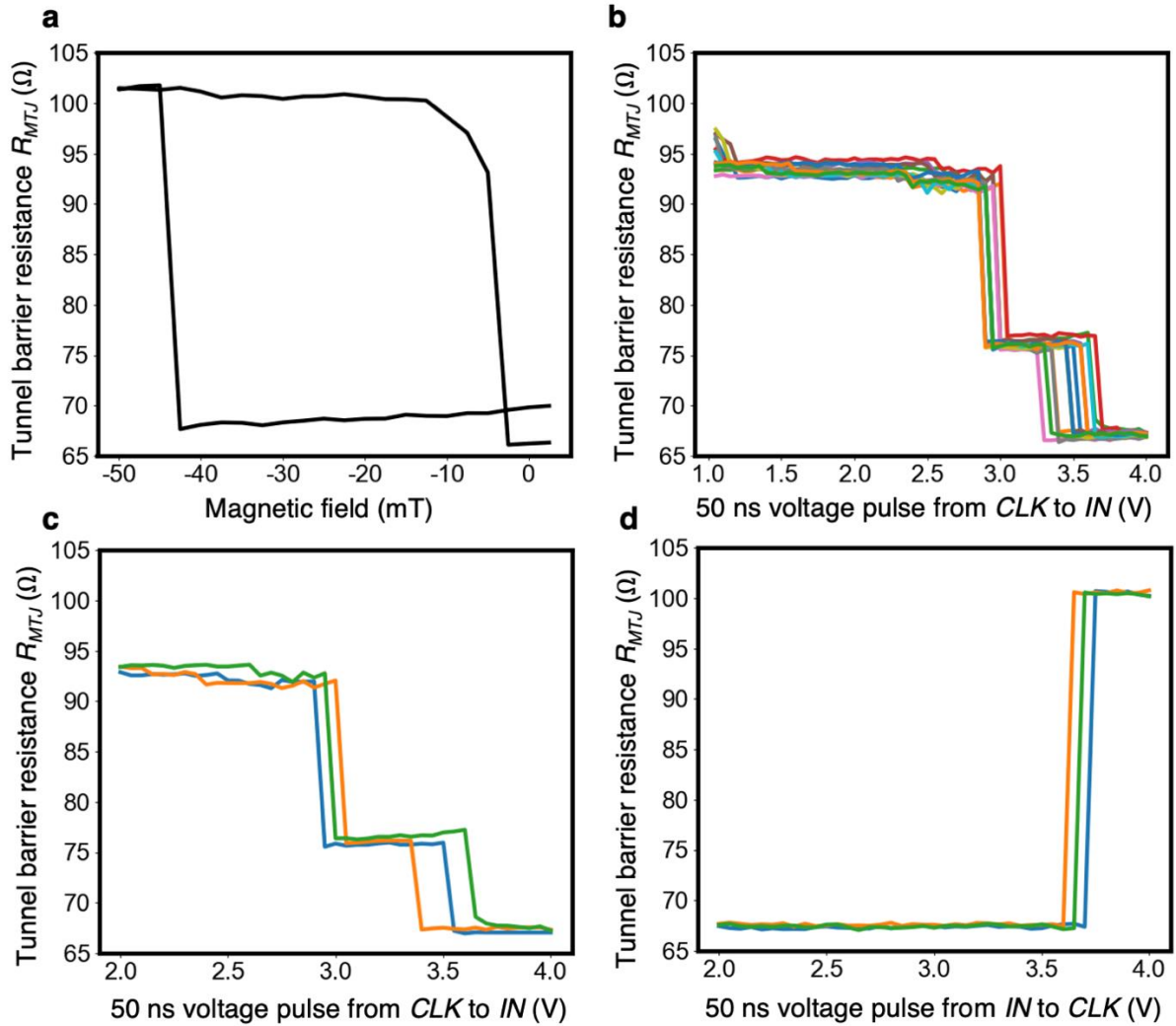

**Figure 2. Trapezoidal synapse behavior. a**, Out-of-plane minor field loop, with the magnetic field applied in $\hat{z}$. **b**, Electrical switching behavior for 13 repeated cycles. **c**, Write and **d**, reset cycling showing MW switching in one direction and single-shot reset behavior after each MW write.

Figure 2b shows $R_{MTJ}$ vs. $V$, repeated from saturation for 13 cycles. A larger track width reduces the lateral current density that impinges on the DW for a given voltage. Therefore, as the DW is moved from *IN* to *CLK*, the voltage amplitude required to propagate the DW increases as the track widens, which results in voltage-controlled MW switching since each notch has a higher depinning voltage than the notch to its left. In all 13 measured cycles the DW reaches the first notch by 1.5 V. The resistance of the first notch is stable at 93.3 Ω ± 1.1 Ω, where the error is the standard deviation in $R_{MTJ}$ over all cycles. The critical switching voltage amplitude $V_C$ for the next switch is, on average, $V_{C,avg} = 3.0\ V$, and defining percent variation $var = \frac{1}{2} * \frac{V_{C,max} - V_{C,min}}{V_{C,avg}}$, this switch has $var = 2.5\%$ over 13 cycles, after which $R_{MTJ} = 76.1\ \Omega \pm 0.4\ \Omega$. The next switch shows $V_{C,avg} = 3.5\ V$ with $var = 5.7\%$, after which $R_{MTJ} = 67.1\ \Omega \pm 0.3\ \Omega$. The trapezoidal geometry and multiple notches reduce the variation in switching voltage measured here compared to our previous work of 10%[20]. See Supplementary Fig. 1 showing data that the MW states are stable over time at room temperature.

Figures 2c-d demonstrate that the nonlinear device geometry gives rise to an asymmetric switching response to the voltage pulse polarity. Here, a similar method is used as in Figure 2b; consecutively higher voltage amplitudes $+V$ are applied to push the DW from *IN* to *CLK* showing multiple weights, and then voltage amplitudes $-V$ are applied showing only a single state transition from the maximum to minimum resistance, as the DW moves from the *CLK* to the *IN* terminal. This is repeated over 3 cycles without resetting the device in-between. The DC bias field is adjusted to promote efficient switching, alternating between -4 mT for MW writing and -42 mT for single-shot resetting. In the write direction there are only three distinct resistance states at 93 Ω, 76 Ω, and 67 Ω; in the single-shot reset direction there are two stable resistance states at 67 Ω and 100 Ω. Utilizing both write and reset operations, four stable and repeatable resistance levels are observed in this device.

In the reset direction, the DW does not exhibit MW switching because depinning from the right-most notch requires the maximum voltage $V_{C,avg} = 3.7\ V$ with $var = 1.35\%$, slightly above the maximum switching voltage in the MW direction of 3.5 $V$. Since MW switching only occurs in one direction, we achieve a dependable reset at the highest voltage amplitude since the DW will not get stuck at one of the intermediate weights during reset.

In Fig. 1b there are seven intermediate notches and therefore there should be 9 achievable resistance states in this device, while only 4 states were resolved. The SEM image shows there was an inconsistency in the fabrication process where the notches had varying geometric depths, and thus they varied in their ability to pin the DW. The deeper notches pinned the DW more effectively, and the shallower notches were passed over. To resolve all MW states, future devices will require better control of the lithographic notch size and geometry, which could be achieved by altering the notch shape[26,27] and using varying doses to more precisely pattern the notches. In Supplementary Fig. 2, we show results of five write/reset cycles of an additional trapezoidal synapse device. Five consistent states are observed in the MW direction and a single switch in the reset direction, but this device had lower TMR.

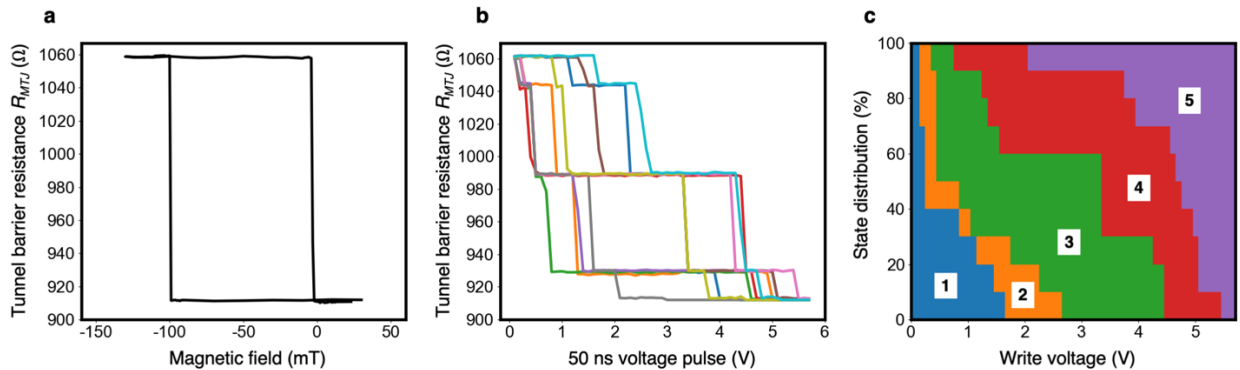

**Figure 3. Straight synapse behavior. a**, Out-of-plane minor field loop, with the magnetic field applied in $\hat{z}$. **b**, MTJ resistance as a function of voltage pulse amplitude across the DW track. The voltage amplitude is increased between 0 and 5.5 V ten times; each color shows the response for one cycle. **c**, State distribution map showing the likelihood for the DW to be in a particular notch at a particular voltage. At a given voltage, the height of a colored region is proportional to the probability that the DW occupies the corresponding notch.

Figure 3 demonstrates results from electrical characterization of the straight synapse depicted in Fig. 1c-d. Figure 3a shows the out-of-plane minor field loop, showing TMR = 16%, RA = 380 $\Omega -$ μm², and a 150 Ω difference between the maximum and minimum MTJ resistances. The reduced TMR compared to the trapezoidal synapse is due to additional resistance at the junction and/or contacts. Nevertheless, perpendicular magnetic anisotropy and a similar offset minor field loop is observed.

The straight synapse's stochastic MW switching behavior is shown in Fig. 3b. Following the same method and parameters as the trapezoidal synapse, the device is first saturated in a DC bias field of 120 mT ($-\hat{z}$) to set the magnetization of the free and reference layers fully antiparallel. After saturation, the DC 4 mT ($-\hat{z}$) bias field is set and 50 ns voltage pulses V are applied between the IN and CLK terminals with increasing amplitude. The DW is re-initialized and the process is repeated over ten cycles. Five resistance levels are observed, showing the DW is pinned by all five notches present. While the stochasticity in $V_C$ is evident over the 10 cycles, the resistance of each MW level is extremely stable, with $R_{MTJ} = 1061.0 \pm 1.1\ \Omega, 1043.3 \pm 2.6\ \Omega, 989.0 \pm 2.1\ \Omega, 929.5 \pm 0.8\ \Omega$, and $913.1 \pm 0.4\ \Omega$ at each respective notch. Here, the depinning energy for each notch is nominally identical, so the final DW position is not a function of the voltage alone and the DW is more liable to move over multiple notches with a single pulse. This process is random; the switching behavior can be statistically analyzed but not accurately predicted, making it potentially useful for stochastic neural networks. This varies significantly from the trapezoidal synapse, which showed repeatable, voltage-controlled MW switching with less variation.

The stochastic behavior is further exemplified in Fig. 3c, where the probability of being in a particular resistance state is plotted against the voltage pulse amplitude, showing five regions that correspond to the five notches. Previous work has investigated constant voltage pulse measurements in binary MTJs[24], where a constant current pulse is sent through the MTJ stack for a fixed number of times, and the probability that the magnetization of the free layer switches is a function of the temperature, with the switching probability emulating a sigmoid. Here, only one current pulse is sent for each voltage, and the voltage is continuously ramped upwards. This allows for a reduction in the number of pulses required, since the bias, and therefore probability, is based on a single voltage pulse. In addition, while variation in weight updates is usually detrimental to neural network accuracy, stochastic updates similar to what is shown in Fig. 3c can offset the negative impact of weight quantization. Small updates in deterministically quantized systems are lost when the update is not large enough to advance the synaptic weight to the next level. Similar to stochastic rounding techniques in software, a notch position switch probabilistically occurs in the DW-MTJ system even with a small update, creating an effectively higher resolution than the number of levels provided by the notches when averaged over thousands of updates, which we have previously shown leads to greater neural network accuracy[7].

The prototypes are 250-350 nm wide with 500 nm-1 μm long MTJs: this size made the nanofabrication more feasible, but also increased the likelihood of MTJ breakdown through pinholes and defects in the relatively large-area MTJ. We have shown, in simulation, scaling to 50 nm widths and comparison of the predicted scaled behavior to other emerging synaptic memory devices[7]. This feature size has been achieved in fabricated MTJ devices[13]. The necessary scaled device length depends on the number of states needed. For example, to obtain 8 states with a minimum feature size of 15 nm (requiring 30 nm separation between notches) would require of the notches and MTJ to span a total length of 240 nm.

In the following sections, we use the device data to inform model parameters and then simulate neuromorphic computing tasks to show the usefulness of the two different geometries: the trapezoidal synapse for stream learning and the linear synapse for inference.

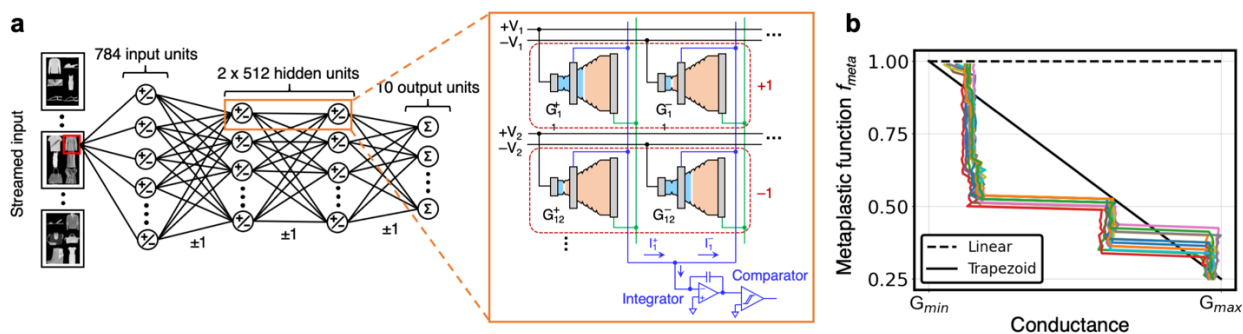

**Figure 4. Metaplastic function using trapezoidal DW-MTJ. a**, Network architecture of ANN applied on the stream Fashion-MNIST task, where circles represent neurons and connecting lines represent DW-MTJ synapses. Activations and binarizations are shown, along with a visual depiction of the stream Fashion-MNIST task. The circuit diagram depicting differential operation of the DW-MTJ devices is shown in the orange box. Blue (green) lines show the current path during inference (training). **b**, Graphical representation of the metaplastic function $f_{meta}$ vs. synaptic weight. The experimental device data expressed in terms of $f_{meta}$ is shown by the colored lines, as well as $f_{meta}$ for the linear synapse (black dashed line) and a trapezoidal DW-MTJ synapse with a 1:4 width ratio (black solid line).

In this section, we show how a trapezoidal DW-MTJ can be optimized for stream learning by taking advantage of tunable nonlinearity. A nonlinear update response, where the amount of conductance change produced by a given pulse depends on the initial conductance, has been shown to have beneficial properties. For example, nonlinear synapses in spiking network arrays using spike-timing dependent plasticity train faster and more accurately than linear synapses[28]. In another example that we will focus on here, binarized NN synapses with nonlinear metaplastic hidden weights are shown to be able to combat catastrophic forgetting and reduce overfitting in low-information situations[29]. In Reference [29], the update is applied to the full analog weight, but the analog weight is binarized according to the sign of the weight to either -1 or +1. As a result, each synapse can be thought of as having an apparent state (binarized) and a metaplastic state (analog) where the metaplastic weight informs the binarized weight. The effect of this binarized metaplasticity is that only accumulated changes that change the sign of the weight will affect network performance. In addition, in Reference [29], a nonlinear and asymmetric update is applied to the weight that introduces a variable learning rate as a function of the analog weight, where updates are larger for weights closer to 0 and smaller for weights with larger magnitudes (nonlinear), but only for updates that prescribe to decrease the weight (asymmetric). The authors then show that the combination of these factors ensures that learning with information-limited datasets has reduced overfitting and therefore better performance.

Here, we use the trapezoidal DW-MTJ to implement such synaptic metaplasticity in hardware. The DW propagating in the track acts as the hidden, metaplastic weight and binarization can be introduced either by patterning a small sensing MTJ that covers only a fraction of the length of the track, or by using a binary comparator circuit if using two devices differentially, shown in the inset circuit in Fig. 4a. In this application, following Reference [29], the applied voltage V will set the DW to multiple states depending on what notch it is set to, but the MTJ itself does not have to produce MW values and could be binary. This could be attractive for the DW-MTJ since the MTJ can be kept small and centered along the DW track. By using a binarized readout while holding a nonlinear multi-state hidden weight, we show that the DW-MTJ can reduce system complexity requirements while providing the capability to prevent overfitting and forgetting of sequentially learned information.

We translate the trapezoidal DW-MTJ MW switching characteristics (Fig. 2b) into a metaplastic function $f_{meta}$ that represents how easy (low voltage required, i.e. DW at the narrow side of the trapezoid with width $w_1$) or hard (high voltage required, i.e. DW at the wide side of the trapezoid with width $w_2$) it is to update the DW position. Thus, $f_{meta}$ is defined such that $f_{meta} = 1$ when the DW is at the narrow end (synaptic weight of 0) and $f_{meta}$ linearly decreases as the DW gets to the wider edge, becoming harder to move (synaptic weight of 1):

(Eq. 1) $$f_{meta} = (1 - |W^H|)\frac{w_2 - w_1}{w_2} + \frac{w_1}{w_2}$$

where $W^H$ is the metaplastic weight represented by the DW position. This function in comparison to the Fig. 2b data is shown in Fig. 4b. The allowed values of the weights are constrained between -1 and 1 by clipping[30] the values at the end of each update to emulate physical limitations of the allowed DW positions. The result is a metaplastic (second-order) synapse where the magnitude of the applied update is a function of the synaptic weight. In the trapezoidal DW-MTJ device studied, there is a 1:4 ratio between $w_1$ and $w_2$, and a corresponding 1:4 ratio in depinning voltage $V_C$, which is included in the model.

To implement the metaplastic update defined in Eq. 1, the algorithm is adapted from Reference [29] and detailed in Supplementary Section 3. The optimizer update is obtained using a form of stochastic gradient descent, and then the metaplastic update is obtained by multiplying the metaplastic function by the optimizer update. For updates that prescribe to increase the magnitude of the weight, an additional asymmetry parameter δ is multiplied to increase the learning rate toward weights with larger values, which can be incorporated into a physical circuit by using larger current sources for positive weight updates.

To show the network performance of this type of nonlinear DW-MTJ synapse, a learning task is performed in which only a subset of the training set is available to the network at a time, as is always the case in online learning. We developed a modified streamed Fashion-MNIST[31] database by splitting the 60,000 images in the training set into 60 subsets of 1000 images each, shown to the network for 30 epochs. The NN is shown in Fig. 4a and detailed in Methods.

To approximate the effect of notches along the track, quantization to 4 levels is applied to the metaplastic weights after each update. Quantization is applied by stochastically rounding the weights to a defined level, mirroring the non-deterministic spread in depinning voltages for each notch seen in Fig. 2b. Supplementary Information Fig. 3 provides results on the difference between stochastic and deterministic rounding.

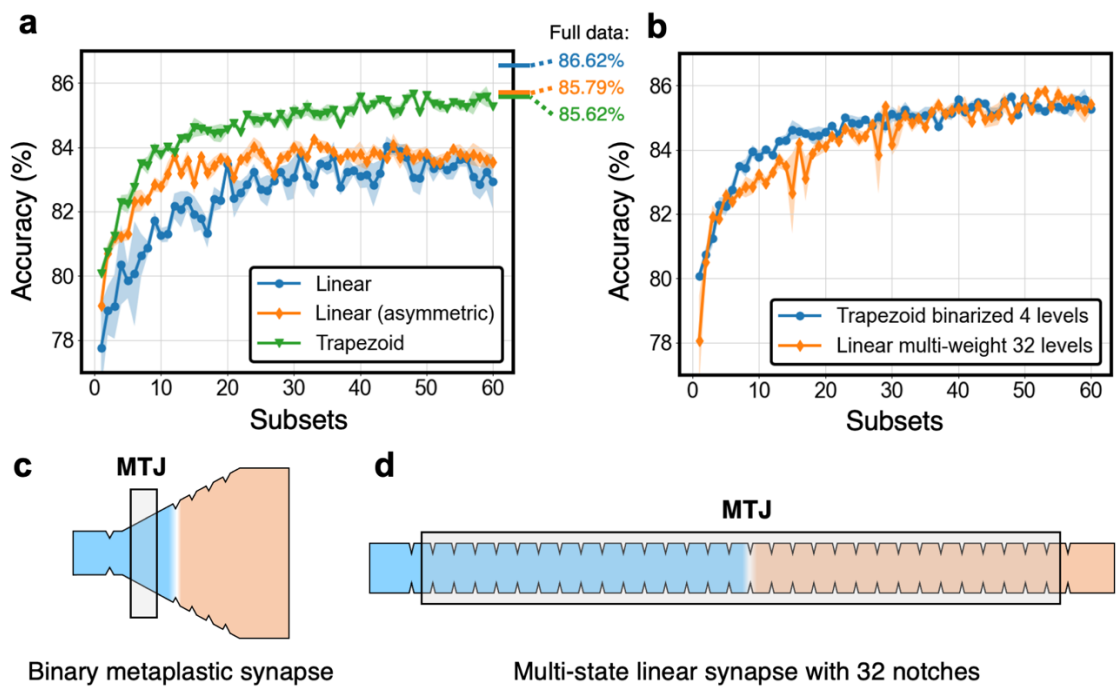

**Figure 5. Trapezoidal DW-MTJ stream learning results. a**, Validation accuracy after 30 epochs as a function of each completed subset of the Fashion-MNIST dataset for different device geometries. Shaded regions are the accuracy distribution due to the 5 different seeds. Results of training with the full dataset for 30 epochs is depicted by the ticks in the top right (top blue = linear, middle orange = linear asymmetric, bottom green = trapezoidal). **b**, Comparison of training performance between the **c**, trapezoidal binarized synapse with the metaplastic weight quantized to 4 levels and **d**, linear multi-weight synapse quantized to 32 levels.

Figure 5a shows the training results of different DW-MTJ synapse geometries, plotting the validation accuracy vs. subset as the different subsets of the data are released to the ANN. Three device types are put through the simulator: DW-MTJ synapses with 1:1 width ratio and asymmetry parameter $\delta = 1$ to represent a straight DW-MTJ synapse with symmetric updates (blue curves with circles); a similar straight DW-MTJ synapse with $\delta = 3$ to control for the effect of asymmetric updates (orange curves with diamonds), and DW-MTJ synapses with 1:4 width ratio and $\delta = 3$ to reflect the trapezoidal synapse data (green curves with triangles). For all three device types, their validation accuracy is also calculated for when they are trained on the full data set for 30 epochs (horizontal ticks in Fig. 5a) to match the number of epochs trained per subset for the streamed data set.

The trapezoidal DW-MTJ synapse (green triangles) provides a clear advantage in the stream Fashion-MNIST task, reaching close to 86% accuracy once all the subsets have been presented, the same as when training the network by presenting all the data at once. The linear synapse reached a final accuracy of around 83% regardless of update asymmetry $\delta$ (blue circles and orange diamonds). There was an advantage for the linear synapse with asymmetric updates ($\delta = 3$, orange diamonds) until around 35 subsets, but this can be attributed to the fact that updates that increase weights have a three times higher learning rate. Supporting Information Fig. 3 provides further results confirming more effective learning for larger trapezoidal width ratios.

In addition, the binarization scheme with stochastic rounding of weights is very resilient to the effects of heavy quantization, shown in Supplementary Fig. 3, where training performance for binarized trapezoidal synapses remain the same for 4, 8 and 16 levels. This is because the effects of quantization are masked due to binarization, and the stochastic weight updates prevent small updates from being lost. In Fig. 5b, the training performance of the trapezoidal binarized synapse with the metaplastic weight quantized to 4 levels is compared to the performance of a non-binarized linear synapse with weight quantization of 32 levels. From the results shown, the final accuracy is roughly the same, approaching 86%. The linear synapse was selected because it performed best when evaluating non-binarized training performance, shown in Supplementary Information Section 5. However, there are clear benefits to using the binarized synapse over a typical non-binarized linear synapse. These benefits are clear when comparing the two devices in Fig. 5c and 5d, where the MTJ size requirement for the binarized synapse is much smaller, reducing the chance for pinholes to occur. The fabrication requirements in terms of patterning the track are lower as well, only requiring patterning of 4 relatively uniform levels as opposed to 32. Due to the level requirement, the linear synapse is also much less compact than that of the trapezoidal binarized synapse.

We see here the advantage of the trapezoidal synapse over a straight synapse for stream learning. For a network consisting of linear synapses, overfitting on the current subset can occur, resulting in the forgetting of details learned in previous subsets. Specifically, training a network for many epochs using a limited dataset will result in an overfitting of the network based on the earliest presented information without effectively consolidating information across streamed batches. Thus, the network loses details learned in previous subsets that may be significant to the dataset overall.

Instead, the trapezoidal metaplastic synapse learns more effectively in the stream learning scenario since there is a weight-dependent learning rate attributed to each update. The trapezoidal metaplastic synapse can retain this information, since large magnitude weights have an effectively lower learning rate compared to weights closer to 0. This allows the network to remember the values of larger weights for a longer period of time while maintaining learning flexibility in synapses with smaller weights. The result is a well-suited application for trapezoidal DW-MTJ synapses with low power, low information needed, and reduced lithographic complexity requirements due to only requiring a binary MTJ and few notches. Moreover, the ability to freely modify the metaplastic function through lithographic definition of the ferromagnetic track to modulate the metaplastic function is another indication that the DW-MTJ synapse is well suited for online learning applications.

INFERENCE TASK USING STRAIGHT MAGNETIC SYNAPSE GEOMETRY

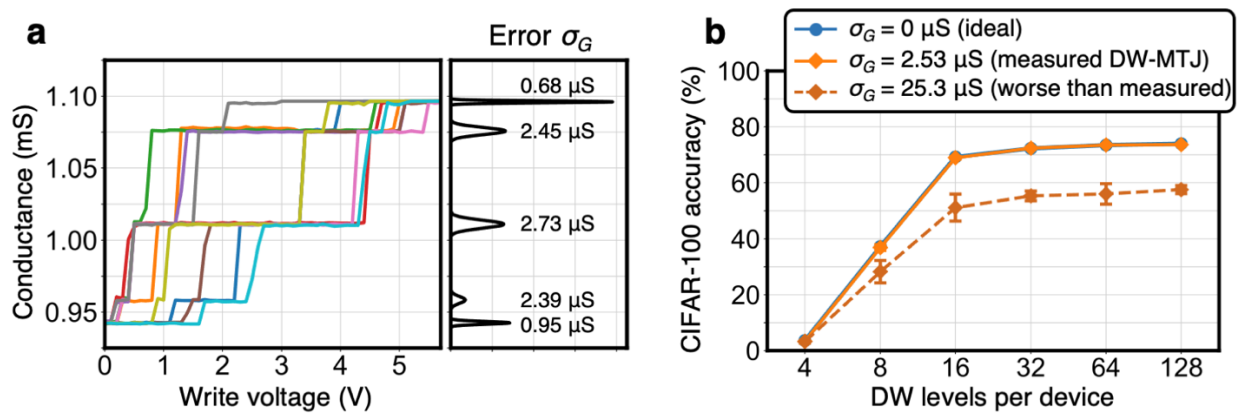

**Figure 6. CIFAR-100 classification task using straight DW-MTJ synapse. a**, Histogram of measured conductances showing five distinct peaks corresponding to the five conductance states that were measured over ten voltage cycles and corresponding with the five notches in the device. **b**, Accuracy on the CIFAR-100 image classification dataset using the ResNet56 network, with no weight errors (blue circles), with weight errors corresponding to the measured DW-MTJ cycle-to-cycle write noise (orange diamonds), and with 10× the measured write noise in the DW-MTJs to show effect of higher variation (orange diamonds, dotted line).

Here, we simulate an inference task that takes advantage of the high stability of the straight DW-MTJ synapse resistance levels and is not drastically affected by the randomness observed in setting the levels. When using the DW-MTJ for an NN inference application, the DW-MTJs are programmed to their desired weight levels, using a write-verify process with multiple pulses per device as needed to overcome the effects of stochasticity. This is a one-time programming operation and is not needed after the system is deployed for the inference application. Figure 6a shows the measured error over 10 cycles in the conductance for each conductance weight, $\sigma_G$; $\sigma_G = 0.68 - 2.73$ μS shows that once the straight DW-MTJ synapse is updated, the conductance levels are highly precise. Cycle-to-cycle write noise leads to a conductance error of only about 0.25% for the notches below the MTJ. This very high precision is a consequence of the fact that the DW is only being pinned at notches whose position cannot move, because they are lithographically defined.

Low write noise allows more precise representation of NN weights. To evaluate the effect on accuracy, we use CrossSim[32] to simulate analog inference in a DW-MTJ system (see Methods for details). For the evaluation, we use the CIFAR-100 classification task[33]: 32×32 pixel images with 3 color channels are classified into one of 100 object categories. This is a significantly more challenging task than MNIST or CIFAR-10, and it requires higher precision in the weights. After mapping weights to conductance, a conductance error is applied that is sampled from a normal distribution with zero mean and a standard deviation of 2.53 µS, corresponding to the measured write noise in Fig. 6a.

Figure 6b shows the CIFAR-100 inference accuracy as a function of the number of available notches in the DW-MTJ. At least 16 notches are needed to obtain a reasonably high accuracy on CIFAR-100; this is a consequence of the NN's intrinsic sensitivity to weight quantization[34]. More importantly, we compare ideal DW-MTJ devices with no conductance error (blue) and DW-MTJ devices with the experimentally characterized level of conductance error (orange). These curves are effectively indistinguishable, showing that the amount of cycle-to-cycle write noise in the measured DW-MTJ devices is low enough to be considered negligible. We note that the conductance error used is based on the properties of a single device and does not include the effect of device-to-device variations. Thus, we also performed the inference task with a write noise that is 10× larger than the measured value, to understand the effect of device-to-device variability. Figure 6b shows that with this higher write noise the accuracy has degraded, but it is still well above zero. This again suggests that the measured level of DW-MTJ write noise is well beyond what is sufficient to obtain high accuracy on CIFAR-100.

## COMPARISON TO OTHER SYNAPSE TECHNOLOGIES

| Device Type | Device | Feature size | Write duration | Write amplitude | Update energy | Linearity | # of states* | Read noise | Write noise |
|---|---|---|---|---|---|---|---|---|---|
| Magnetic | DW-MTJ (this work) | 0.35 × 3.6 µm² (50 nm notch) | 50 ns | +0.2/-0.2 V | 0.124 pJ | Tunable | 5 | Very low (0.25%) | - |
| | DW-MTJ (scaled)[7] | 50 × 1000 nm² (5 nm notch) | 0.5 ns | +27/-27 µA | 2.5 fJ | High (+0.80/-0.81) | >32 | - | Low (3.23%) |
| | 2T-MTJ[25] | $\frac{\pi}{4}$ × 100 × 40 nm² | 5 ns | +1/-1 V | 38 fJ | N/A | 2 | - | Low |
| Resistive | TaO$_x$/HfO$_x$[35,36] | ~1 µm² | 50 ns | +1.6/-1.6 V | 1.28 pJ | High (+0.04/-0.63) | 128 | High | Low (3.7%) |
| | TaO$_x$[37] | ~1 µm² | 10 ns | +1/-1 V | ~10 nJ | Low (+668/-51.7) | - | High | High (11.20%) |
| Phase Change | hBEC-PCM (GST)[38] | 300 × ~1000 nm² | 20 ns | +1.25/+1.6 V | ~0.5 pJ | Low (+3.00/-1.00) | 200/30 | High | High (~15%) |
| Ferroelectric | HfZrO$_x$[39] | 1,190.4 µm² | 75 ns | +3.65/-2.95 V | 1.79 pJ | Med (+1.75/+1.46) | 32 | Med | Low (0.5%) |
| Electrochemical | PEDOT:PSS/EMIM:TFSI PVDF-HFP[40,41] | 750 × 2000 µm² | 250 µs (1 µs settling) | +1/-1 V | 4 fJ µm⁻² | High (+0.70/-0.12) | >100 | Low | Low (0.023%) |
| | p(g2T-TT)/EMIM:TFSI PVDF-HFP[40] | 30 × 30 µm² | 20 ns (100 ns settling) | +1/-1 V | 10 fJ µm⁻² | High | >100 | Low | Low |
| Charge Trap | SONOS[42] | 7 × 1.2 µm² | 10 µs | +10/-11 V | 8.2 fJ µm⁻² | Med (+1.59/-2.22) | ~50 | Low | Low (0.23%) |

**Table 1. Performance comparison of a variety of synaptic devices.** *Number of states is included for completeness, but the actual achievable number of states depends on read and write noise.

In this section, the DW-MTJ is compared against state-of-the-art alternatives for artificial synapses. The stand-out feature of DW-MTJ synapses is their relatively high speed and low update energy. Though an update costs around 0.124 pJ in this work, it has been demonstrated in our previous simulation work that this design can be scaled down, leading to lower current densities and smaller update energies. Additionally, though the number of states for the DW-MTJ in this work is five, this can be increased by adding more notches to define more levels, if needed. The DW-MTJ has tunable nonlinearity, which enables unique neuromorphic functionality like the metaplastic function. Electrochemical devices, in particular, show extreme promise with very low write noise to hold many states, as well as extremely linear and symmetric behavior. However, these devices operate in much slower timescales than the operation speed of the DW-MTJ device. These advantages set up DW-MTJ artificial synapses as a strong candidate for online learning applications.

CONCLUSIONS

In conclusion, we have shown DW-MTJ devices can achieve multiple weights with low-noise resistance levels, and that a single materials stack can yield devices with complex multi-functionality that is tunable via patterning geometry. Using neuromorphic models based on experimental data, we showed, as an example, that the trapezoidal geometry can create a metaplastic function useful for online learning in information-limited environments, where the NN has to learn as the data arrives. We also showed that the extremely low read noise in the straight geometry produces excellent inference performance when compared to other synaptic devices, approaching software accuracy for quantized systems. This work shows MW magnetic synapses are a feasible technology for neuromorphic computing and provides design guidelines for emerging synapse technologies.

METHODS

**Fabrication of DW-MTJ synapse devices**

The device fabrication consisted of eight main steps starting with the thin film stack. Patterning of all features was performed using an E-line Raith electron beam lithography (EBL) tool using negative-tone Ma-N 2405 and positive-tone PMMA-A4 resists. The first step was to pattern and then etch the DW track using an AJA International ion miller. EBL was used to pattern the MTJ above the DW track and then etched with the ion miller to the MgO layer. Vias were patterned and silicon nitride was deposited using a Plasma-Therm plasma-enhanced chemical vapor deposition (PECVD) tool. After encapsulation, Cr/Au (5/95 nm) contacts were deposited with a Kurt J. Lesker PVD75 e-beam evaporator.

**Electrical testing of DW-MTJ synapse devices**

A West Bond wire bonder was used to connect the devices to a home-built testing setup with a tunable electromagnet. Electrical testing was performed by saturating the devices in an out-of-plane magnetic field to achieve the desired configuration, parallel or antiparallel and then set to a DC bias field as described in the main text. 50 ns (with 10 ns rise and fall times) voltage pulses were used to move the DW along the track and the 4-point resistance of the MTJ was measured after each pulse.

**Setup of stream Fashion-MNIST training task**

The Fashion-MNIST[31] database of 60,000 images in the training set is split into 60 subsets of 1000 images each, shown to the network for 30 epochs. In between every epoch, the network's performance is quantified as it performs inference on a separate validation set of 10,000 images. The network is trained sequentially on subsets for 30 epochs each, and all 60 subsets are trained for 5 different seeds (i.e. 5 different randomly initialized weights). The network architecture is a fully connected binarized NN with 784 input units, two sign activation hidden layers with 512 units each, and 10 output units, with batch normalization applied after each layer. The Adam optimizer[43] is used with a learning rate of 0.005 for all runs. The network is implemented in PyTorch[44].

**Setup of CIFAR-10 inference task**

A NN is first trained using Keras[45] over 200 epochs without accounting for any physical non-idealities. We use the ResNet56 topology, which has the same structure as the identically named network in Ref. [10] for CIFAR-10, but has 4× more channels in every convolutional layer to accommodate the greater complexity of the CIFAR-100 dataset. The final network obtains a software accuracy of 73.9% on the test set and contains 13.7M floating-point weights, which must be quantized to $N$ bits before mapping to hardware. In CrossSim, an $N$-bit signed weight is mapped to the difference in conductance of two DW-MTJ devices each: if the weight is positive, one device encodes the weight's magnitude, while the other device is set to the minimum conductance (0.94 mS), and vice versa if the weight is negative. Each device has $2^{N-1}$ distinct conductance levels between 0.94 mS and 1.1 mS, implemented using $2^{N-1}$ DW notches. A larger $N$ increases the available precision, but requires more notches and a longer DW-MTJ device.

## ASSOCIATED CONTENT

*Supplementary Information* is available and includes resistance state stability data, data on an additional trapezoidal device, and additional data from the stream learning task.

## AUTHOR INFORMATION

*Corresponding Author*

*Email: incorvia@austin.utexas.edu

*Email: cbennet@sandia.gov*Author Contributions*

T. L. fabricated and measured the devices assisted by C. C. and O. G. A. M. A. measured the devices. S. L., T. P. X., and C. H. B. generated and carried out the neural network simulations. L. X. grew the film stack. J. S. F. provided technical guidance. M. J. M. led the Sandia-based team. M. J. M. and C. H. B. conceived the idea for the trapezoidal synapse. J. A. C. I. led the UT-Austin-based team and the project. J. A. C. I., T. L., and S. L. wrote the manuscript with input from all authors. All authors discussed the results.

*Acknowledgments*

This research is sponsored in part by the National Science Foundation under CCF awards 1910997 and 1910800. The authors acknowledge funding from the National Science Foundation CAREER under award number 1940788. This material is also based upon work supported by the National Science Foundation Graduate Research Fellowship under Grant Nos. 2020307514 and 2021311125. This research is sponsored in part by Sandia's Laboratory-Directed Research and Development program. This paper describes objective technical results and analysis. Any subjective views or opinions that might be expressed in this paper do not necessarily represent the views of the U.S. Department of Energy or the United States Government. Sandia National Laboratories is a multimission laboratory managed and operated by NTESS, LLC, a wholly owned subsidiary of Honeywell International Inc., for the U.S. Department of Energy's National Nuclear Security Administration under Contract No. DE-NA0003525.

The work was done at the Texas Nanofabrication Facility supported by NSF Grant No. NNCI-1542159 and at the Texas Materials Institute (TMI). The authors acknowledge computing resources from the Texas Advanced Computing Center (TACC) at the University of Texas at Austin (http://www.tacc.utexas.edu).

*Notes*

The authors declare no competing financial interest.

REFERENCES

1. B. Y. J. B. Aimone. review articles Algorithms and Computing Beyond Moore ' s Law.

2. F. Stella, P. Baracskay, J. O'Neill & J. Csicsvari. Hippocampal Reactivation of Random Trajectories Resembling Brownian Diffusion. *Neuron* **102,** 450-461.e7 (2019).

3. U. Kraushaar & P. Jonas. Efficacy and stability of quantal GABA release at a hippocampal interneuron-principal neuron synapse. *Journal of Neuroscience* **20,** 5594–5607 (2000).

4. D. A. Henze, N. N. Urban & G. Barrionuevo. The multifarious hippocampal mossy fiber pathway: A review. *Neuroscience* **98,** 407–427 (2000).

5. D. Pecevski, L. Buesing & W. Maass. Probabilistic inference in general graphical models through sampling in stochastic networks of spiking neurons. *PLoS Computational Biology* **7,** (2011).

6. Y. Shim, S. Chen, A. Sengupta & K. Roy. Stochastic Spin-Orbit Torque Devices as Elements for Bayesian Inference. *Scientific Reports* **7,** 14101 (2017).

7. S. Liu, T. P. Xiao, C. Cui, J. A. C. Incorvia, C. H. Bennett & M. J. Marinella. A domain wall-magnetic tunnel junction artificial synapse with notched geometry for accurate and efficient training of deep neural networks. *Applied Physics Letters* **118,** 202405 (2021).

8. M. Fränzl & F. Cichos. Active particle feedback control with a single-shot detection convolutional neural network. *Scientific Reports* **10,** 1–7 (2020).

9. S. Park, J. Noh, M. L. Choo, A. M. Sheri, M. Chang, Y. B. Kim, C. J. Kim, M. Jeon, B. G. Lee, B. H. Lee & H. Hwang. Nanoscale RRAM-based synaptic electronics: Toward a neuromorphic computing device. *Nanotechnology* **24,** 0–6 (2013).

10. K. He, X. Zhang, S. Ren & J. Sun. Deep residual learning for image recognition. *Proceedings of the IEEE Computer Society Conference on Computer Vision and Pattern Recognition* **2016-December,** 770–778 (2016).

11. B. Sutton, K. Y. Camsari, B. Behin-Aein & S. Datta. Intrinsic optimization using stochastic nanomagnets. *Scientific Reports* **7,** 1–10 (2017).

12. Z. Chen, B. Gao, Z. Zhou, P. Huang, H. Li, W. Ma, D. Zhu, L. Liu, X. Liu, J. Kang & H. Y. Chen. Optimized learning scheme for grayscale image recognition in a RRAM based analog neuromorphic system. *Technical Digest - International Electron Devices Meeting, IEDM* **2016-Febru,** 17.7.1-17.7.4 (2015).

13. L. Xue, C. Ching, A. Kontos, J. Ahn, X. Wang, R. Whig, H. W. Tseng, J. Howarth, S. Hassan, H. Chen, M. Bangar, S. Liang, R. Wang & M. Pakala. Process optimization of perpendicular magnetic tunnel junction arrays for last-level cache beyond 7 nm node. *Digest of Technical Papers - Symposium on VLSI Technology* **2018-June,** 117–118 (2018).

14. A. Sengupta, Y. Shim & K. Roy. Proposal for an all-spin artificial neural network: Emulating neural and synaptic functionalities through domain wall motion in ferromagnets. *IEEE Transactions on Biomedical Circuits and Systems* **10,** 1152–1160 (2016).

15. O. Akinola, X. Hu, C. H. Bennett, M. Marinella, J. S. Friedman & J. A. C. Incorvia. Three-terminal magnetic tunnel junction synapse circuits showing spike-timing-dependent plasticity. *Journal of Physics D: Applied Physics* **52,** 49LT01 (2019).

16. N. Hassan, X. Hu, L. Jiang-Wei, W. H. Brigner, O. G. Akinola, F. Garcia-Sanchez, M. Pasquale, C. H. Bennett, J. A. C. Incorvia & J. S. Friedman. Magnetic domain wall neuron with lateral inhibition. *Journal of Applied Physics* **124,** (2018).

17. W. H. Brigner, X. Hu, N. Hassan, C. H. Bennett, J. A. C. Incorvia, F. Garcia-Sanchez & J. S. Friedman. Graded-Anisotropy-Induced Magnetic Domain Wall Drift for an Artificial Spintronic Leaky Integrate-and-Fire Neuron. *IEEE Journal on Exploratory Solid-State Computational Devices and Circuits* **5,** 1–1 (2019).

18. S. Liu, C. H. Bennett, J. S. Friedman, M. J. Marinella, D. Paydarfar & J. A. Incorvia. Controllable reset behavior in domain wall-magnetic tunnel junction artificial neurons for task-adaptable computation. *IEEE Magnetics Letters* **11,** 1–5 (2021).

19. C. Cui, O. G. Akinola, N. Hassan, C. H. Bennett, M. J. Marinella, J. S. Friedman & J. A. C. Incorvia. Maximized lateral inhibition in paired magnetic domain wall racetracks for neuromorphic computing. *Nanotechnology* **31,** (2020).

20. M. Alamdar, T. Leonard, C. Cui, B. P. Rimal, L. Xue, O. G. Akinola, T. Patrick Xiao, J. S. Friedman, C. H. Bennett, M. J. Marinella & J. A. C. Incorvia. Domain wall-magnetic tunnel junction spin–orbit torque devices and circuits for in-memory computing. *Applied Physics Letters* **118,** 112401 (2021).

21. E. Raymenants, A. Vaysset, D. Wan, J. Swerts, S. Van Beek, O. Zografos, D. E. Nikonov, S. Manipatruni, I. A. Young, D. Mocuta, I. P. Radu, M. Heyns & M. Manfrini. Spin-torque-driven MTJs with extended free layer for logic applications. *Journal of Physics D: Applied Physics* **51,** (2018).

22. S. A. Siddiqui, S. Dutta, A. Tang, L. Liu, C. A. Ross & M. A. Baldo. Magnetic Domain Wall Based Synaptic and Activation Function Generator for Neuromorphic Accelerators. *Nano Letters* **20,** 1033–1040 (2020).

23. A. F. Vincent, J. Larroque, N. Locatelli, N. Ben Romdhane, O. Bichler, C. Gamrat, W. S. Zhao, J.-O. Klein, S. Galdin-Retailleau & D. Querlioz. Spin-transfer torque magnetic memory as a stochastic memristive synapse for neuromorphic systems. *IEEE Trans Biomed Circuits Syst* **9,** 166–174 (2015).

24. A. Sengupta, P. Panda, P. Wijesinghe, Y. Kim & K. Roy. Magnetic Tunnel Junction Mimics Stochastic Cortical Spiking Neurons. *Scientific Reports* **6,** 30039 (2016).

25. G. Srinivasan, A. Sengupta & K. Roy. Magnetic Tunnel Junction Based Long-Term Short-Term Stochastic Synapse for a Spiking Neural Network with On-Chip STDP Learning. *Scientific Reports* **6,** 29545 (2016).

26. D. Djuhana, H. G. Piao, S. H. Lee, D. H. Kim, S. M. Ahn & S. B. Choe. Asymmetric ground state spin configuration of transverse domain wall on symmetrically notched ferromagnetic nanowires. *Applied Physics Letters* **97,** 95–98 (2010).

27. V. D. Nguyen, W. S. Torres, P. Laczkowski, A. Marty, M. Jamet, C. Beigné, L. Notin, L. Vila & J. P. Attané. Elementary depinning processes of magnetic domain walls under fields and currents. *Scientific Reports* **4,** 1–6 (2014).

28. D. Querlioz, O. Bichler, A. F. Vincent & C. Gamrat. Bioinspired Programming of Memory Devices for Implementing an Inference Engine. *PROCEEDINGS- IEEE* **103,** 1398–1416 (2015).

29. A. Laborieux, M. Ernoult, T. Hirtzlin & D. Querlioz. Synaptic metaplasticity in binarized neural networks. *Nature Communications* **12,** 1–12 (2021).

30. P. Merolla, R. Appuswamy, J. Arthur, S. K. Esser & D. Modha. Deep neural networks are robust to weight binarization and other non-linear distortions. (2016).

31. H. Xiao, K. Rasul & R. Vollgraf. Fashion-MNIST: a Novel Image Dataset for Benchmarking Machine Learning Algorithms. (2017).

32. S. Agarwal, R. B. J. Gedrim, A. H. Hsia, D. R. Hughart, E. J. Fuller, A. A. Talin, C. D. James, S. J. Plimpton & M. J. Marinella. Achieving ideal accuracies in analog neuromorphic computing using periodic carry. *Digest of Technical Papers - Symposium on VLSI Technology* T174–T175 (2017). doi:10.23919/VLSIT.2017.7998164

33. CIFAR-100 Python. *https://www.kaggle.com/fedesoriano/cifar100* Available at: https://www.kaggle.com/fedesoriano/cifar100.

34. S. Gupta, A. Agrawal, K. Gopalakrishnan & P. Narayanan. Deep Learning with Limited Numerical Precision. *Proceedings of the 32nd International Conference on International Conference on Machine Learning - Volume 37* 1737–1746 (2015).

35. W. Wu, H. Wu, B. Gao, P. Yao, X. Zhang, X. Peng, S. Yu & H. Qian. A Methodology to Improve Linearity of Analog RRAM for Neuromorphic Computing. in *2018 IEEE Symposium on VLSI Technology* 103–104 (IEEE, 2018). doi:10.1109/VLSIT.2018.8510690

36. X. Sun & S. Yu. Impact of Non-Ideal Characteristics of Resistive Synaptic Devices on Implementing Convolutional Neural Networks. *IEEE Journal on Emerging and Selected Topics in Circuits and Systems* **9,** 570–579 (2019).

37. C. H. Bennett, D. Garland, R. B. Jacobs-Gedrim, S. Agarwal & M. J. Marinella. Wafer-Scale TaO$_x$ Device Variability and Implications for Neuromorphic Computing Applications. in *2019 IEEE International Reliability Physics Symposium (IRPS)* 1–4 (IEEE, 2019). doi:10.1109/IRPS.2019.8720596

38. S. la Barbera, D. R. B. Ly, G. Navarro, N. Castellani, O. Cueto, G. Bourgeois, B. de Salvo, E. Nowak, D. Querlioz & E. Vianello. Narrow Heater Bottom Electrode-Based Phase Change Memory as a Bidirectional Artificial Synapse. *Advanced Electronic Materials* **4,** 1800223 (2018).

39. M. Jerry, P.-Y. Chen, J. Zhang, P. Sharma, K. Ni, S. Yu & S. Datta. Ferroelectric FET analog synapse for acceleration of deep neural network training. in *2017 IEEE International Electron Devices Meeting (IEDM)* 6.2.1-6.2.4 (IEEE, 2017). doi:10.1109/IEDM.2017.8268338

40. A. Melianas, T. J. Quill, G. LeCroy, Y. Tuchman, H. v. Loo, S. T. Keene, A. Giovannitti, H. R. Lee, I. P. Maria, I. McCulloch & A. Salleo. Temperature-resilient solid-state organic artificial synapses for neuromorphic computing. *Science Advances* **6,** (2020).

41. Y. Li, T. P. Xiao, C. H. Bennett, E. Isele, A. Melianas, H. Tao, M. J. Marinella, A. Salleo, E. J. Fuller & A. A. Talin. In situ Parallel Training of Analog Neural Network Using Electrochemical Random-Access Memory. *Frontiers in Neuroscience* **15,** (2021).

42. S. Agarwal, D. Garland, J. Niroula, R. B. Jacobs-Gedrim, A. Hsia, M. S. van Heukelom, E. Fuller, B. Draper & M. J. Marinella. Using Floating-Gate Memory to Train Ideal Accuracy Neural Networks. *IEEE Journal on Exploratory Solid-State Computational Devices and Circuits* **5,** 52–57 (2019).

43. D. P. Kingma & J. Ba. Adam: A Method for Stochastic Optimization. (2014).

44. A. Paszke, S. Gross, F. Massa, A. Lerer, J. Bradbury, G. Chanan, T. Killeen, Z. Lin, N. Gimelshein, L. Antiga, A. Desmaison, A. Kopf, E. Yang, Z. DeVito, M. Raison, A. Tejani, S. Chilamkurthy, B. Steiner, L. Fang, *et al.* PyTorch: An Imperative Style, High-Performance Deep Learning Library. in *Advances in Neural Information Processing Systems* (eds. Wallach, H., Larochelle, H., Beygelzimer, A., d Alché-Buc, F., Fox, E. & Garnett, R.) **32,** (Curran Associates, Inc., 2019).

45. F. Chollet. Keras. *https://keras.io/* 4 Available at: https://keras.io/.